\documentclass[12pt,showpacs,superscriptaddress,nofootinbib]{revtex4}
\usepackage{amsmath,amssymb,amsthm,amsfonts}
\usepackage{hyperref}

\begin{document}

\title{A brief introduction to Loop Quantum Cosmology}


\author{Guillermo A. Mena Marug\'an}\email{mena@iem.cfmac.csic.es}
\affiliation{Instituto de Estructura de la Materia,
CSIC, Serrano 121, 28006 Madrid, Spain}

\begin{abstract}
In recent years, Loop Quantum Gravity has emerged as a
solid candidate for a nonperturbative quantum theory
of General Relativity. It is a background independent
theory based on a description of the gravitational
field in terms of holonomies and fluxes. In order to
discuss its physical implications, a lot of attention
has been paid to the application of the quantization
techniques of Loop Quantum Gravity to symmetry reduced
models with cosmological solutions, a line of research
that has been called Loop Quantum Cosmology. We
summarize its fundamentals and the main differences
with respect to the more conventional quantization
approaches employed in cosmology until now. In
addition, we comment on the most important results
that have been obtained in Loop Quantum Cosmology by
analyzing simple homogeneous and isotropic models.
These results include the resolution of the classical
big-bang singularity, which is replaced by a quantum
bounce.
\end{abstract}

\maketitle

%%%%%%%%%%%%%%%%%%%%%%%%%%%%%%%%%%%%%%%%%%%%
%% MAINMATTER
%%%%%%%%%%%%%%%%%%%%%%%%%%%%%%%%%%%%%%%%%%%%

\section{Motivation}

Gravity is the only fundamental physical interaction
which is not yet satisfactorily described quantum
mechanically. Even without adhering to the belief that
all fundamental interactions should finally be unified
in a single theory, a strong motivation to search for
a quantum theory of gravity comes from the very own
results of General Relativity. The classical
singularity theorems that arise in Einstein theory
\cite{hawk} imply that (in a variety of physically
relevant situations) the predictability breaks down,
so that the regime of applicability of General
Relativity has been surpassed. Therefore, a new and
more fundamental theory is needed for a correct
physical description.

In trying to quantize General Relativity, the first
obstacle that one finds is that Einstein theory is not
renormalizable as a quantum field theory, so that a
conventional perturbative quantization cannot be
performed. In this context, an alternative
quantization program, known as Loop Quantum Gravity
(LQG), has recently been proposed for General
Relativity \cite{LQG,LQG2,thie}. LQG is an attempt to
construct a nonperturbative quantum theory of gravity
using techniques similar to those of gauge field
theories (e.g., Yang-Mills). The application of these
nonperturbative quantization techniques to simple
gravitational models with application in cosmology,
such as homogeneous and isotropic spacetimes with
different types of matter content, has given rise to a
new branch of gravitational physics called Loop
Quantum Cosmology (LQC) \cite{LQC}.

\section{Hamiltonian formulation of general relativity and
Ashtekar variables}

LQG is a nonperturbative canonical quantization of
General Relativity; therefore, it is constructed
starting from a Hamiltonian formulation of Einstein
theory \cite{ham}. Let us review very briefly this
formulation.

We consider globally hyperbolic four-dimensional
spacetimes $(g_{\alpha\beta}, M = I\!\!\!R \times
\Sigma)$, where $g_{\alpha\beta}$ is a Lorentzian
metric, Greek indices are spacetime indices, and
$\Sigma$ is a three-dimensional manifold. For General
Relativity, once $\Sigma$ is given, the physically
relevant information to determine the classical
solutions is contained in the spatial three-metric
$h_{ab}$ induced on $\Sigma$, and in the corresponding
extrinsic curvature $K_{ab}=\frac{1}{2} {\cal L}_n
h_{ab}$, where Latin indices from the beginning of the
alphabet denote spatial indices, $n$ is the unit
normal to $\Sigma$, and ${\cal L}$ is the Lie
derivative. Equivalently, we can adopt co-triads
$e_a^i$ (rather than spatial metrics) to describe the
system. This allows the coupling to fermionic matter
fields. With $\eta_{ij}$ being the Euclidean
three-metric, we have the relations
\begin{equation} h_{ab}=e_a^i \eta_{ij} e_b^j,\quad
\quad K_a^i=K_{ab}e^b_j \eta^{ij}.\end{equation} Here,
the triad $e^b_j$ is the inverse of the co-triad,
$e_a^i e^b_i=\delta_a^b$ and $e_a^i e^a_j=\delta_j^i$,
and Latin indices from the middle of the alphabet are
internal SU(2) indices, corresponding to the
symmetries of the Euclidean metric $\eta_{ij}$ under
linear transformations.

A set of canonical variables for General Relativity
(in the sense that their Poisson bracket is
proportional to the identity) is given then by the
densitized triad $E^a_i$ and the extrinsic curvature
in triadic form $K_a^i$:
\begin{equation}
E^a_i:=\sqrt{ {\rm det} h}\,e^a_i, \;\; K_a^i\quad
\rightarrow \quad \left\{ K_a^i(x),E^b_j(y)\right\} =
\delta_a^b \delta^i_j \delta^{(3)}(x-y).\end{equation}
In this expression, $x$ and $y$ are two generic points
in $\Sigma$, and we have chosen units such that $8\pi
G =1$, where $G$ is Newton constant.\footnote{In the
following, we also set $\hbar=c=1$.}

Actually, we can replace the triadic extrinsic
curvature by a connection valued 1-form taking values
on su(2), and still obtain a canonical set of
variables. For this, it suffices to realize that the
co-triad determines an su(2)-connection compatible
with it, $\Gamma_a^i$, and notice that the sum of this
connection with any vector (both from the internal and
spatial viewpoints) provides again an su(2)-connection
valued 1-form. Therefore, at the classical level, we
can simply replace $K_a^i$ with
$A^{(\gamma)}\,_a^i=\Gamma_a^i+ \gamma K_a^i$. Here,
$\gamma$ is a nonzero constant called the Immirzi
parameter, and its presence can be seen to lead to an
ambiguity in the quantization \cite{immirzi,immirzi2}
which is usually resolved in LQG by appealing to the
recovery of the Bekenstein-Hawking law for the entropy
of black holes \cite{bhentropy}. For simplicity, we
will set it equal to one from now on. The calculations
for general $\gamma$ can be easily reproduced along
the lines explained below.

We will thus adopt as canonical variables the set
formed by $A_a^i=\Gamma_a^i + K_a^i$ and $E^a_i$. In
General Relativity, these variables are subject to
three types of constraints \cite{thie,ham}. First,
there is a Gauss constraint which generates
SU(2)-transformations,
\begin{equation} {\cal G} _i:=
\partial_a E^a_i + \epsilon_{ij}^{\;\;\;k} A_a^j
E^a_k=0.\end{equation} In addition, the invariance of
the theory under spatial diffeomorphisms is reflected
in the so-called vector or diffeomorphism constraint,
\begin{equation} {\cal V}_a:= F_{ab}^i E^b_i=0,\end{equation}
where $F_{ab}^i$ is the curvature of the connection
$A_a^i$, namely
\begin{equation} F_{ab}^i=2
\partial_{[ a } A_{b ]}^i +\epsilon^i_{\;jk} A_a^j
A_b^k .\end{equation} Here, $\epsilon_{ijk}$ is the
totally antisymmetric symbol. Finally, the invariance
of General Relativity under time reparametrizations
leads to a scalar constraint, also called Hamiltonian
constraint, which in vacuo takes the expression
\begin{equation} {\cal S}:= E^a_i E^b_j \left(
\epsilon_{\;\;\;k}^{ij} F_{ab}^k - 4 K_{[ a}^i
K_{b]}^j \right)=0.\end{equation} Given the
four-dimensional covariance of Einstein theory,
General Relativity is a completely constrained system,
i.e., the total Hamiltonian which generates the
dynamics is just a(n integrated) linear combination of
constraints. In particular, apart from boundary terms,
the Hamiltonian vanishes on classical solutions. On
the other hand, it is worth pointing out that General
Relativity is formulated in terms of connections and
densitized triads without introducing any metric
background structure. This background independence
plays a fundamental role in the theory and will be a
basic guideline for the selection of a quantization
procedure in the construction of LQG.

\section{Holonomy and flux algebra}

Since SU(2)-transformations are symmetries of our
gravitational systems, only the gauge invariant
information about the connection is physically
relevant. Taking into account that this information is
captured by the Wilson loops \cite{referee,referee2},
we can then replace the connection by
SU(2)-holonomies. More specifically, from now on we
will consider holonomies along piecewise
analytic\footnote{The restriction of piecewise
analyticity ensures that the intersection between
edges, as well as the intersection of an edge with a
(piecewise analytic) surface, occurs in a finite
number of points \cite{ashlew}.} edges $e$, where we
understand that an edge is an embedding of the
interval [0,1] in $\Sigma$ \cite{thie,ashlew}. We call
$h_e$ the corresponding holonomy,
\begin{equation}
h_e= {\cal P} \exp{\int_e A_a^i \tau_i
dx^a}.\end{equation} Here, the symbol ${\cal P}$
denotes path ordering, and $\{ \tau_j=-\frac{i}{2}
\sigma_j;\; j=1,2,3\}$ is a basis in the algebra
su(2), with $\sigma_j$ being the Pauli matrices. Let
us notice that the line integral appearing in the
holonomies implies a one-dimensional smearing of the
connection, and that no use of background structures
has been made in the definition of the holonomy.

Since the most relevant field divergences in our
theory are expected to come from the appearance of the
three-dimensional delta function in the basic Poisson
brackets between our variables, and we have already
smeared the connection over one dimension, it seems
natural to try to smear now $E^a_i$ over two
dimensions. Once again, we want to carry out this
smearing without employing any background structure.
Remarkably, this requirement can be fulfilled because
$E^a_i$ is a vector density. Hence, for any piecewise
analytic surface $S$ and any su(2)-valued smooth
function $f^i$ on it, we introduce the associated flux
of the densitized triad,
\begin{equation}
E(S,f)= \int_S E^a_i f^i \epsilon_{abc} \, dx^b dx^c
.\end{equation}

The defined holonomies and fluxes form an algebra
under Poisson brackets. In the following, we take this
algebra as our algebra of elementary phase space
variables. From this perspective, the quantization of
the system amounts to constructing a representation of
this algebra. A keystone result in LQG is a uniqueness
representation theorem known as the LOST theorem
(after the initials of its authors \cite{lost}). The
LOST theorem states that there exists only one cyclic
representation of the holonomy-flux algebra with a
diffeomorphism-invariant state (interpretable as a
``vacuum''). Therefore, the choice of the algebra of
elementary variables, motivated by background
independence, together with the identification of
diffeomorphism invariance as a fundamental symmetry
suffice to pick out a unique quantization (up to
unitary equivalence).

In order to gain insight into the kind of quantization
adopted in LQG, let us first call cylindrical those
complex functions of the connection that depend on it
only via the holonomies along a finite number of edges
(forming a graph \cite{ashlew}). We can identify the
commutative unital $*$-algebra of these functions as
the algebra of configuration variables. By completing
it with respect to the sup-norm\footnote{The use of
this norm is motivated by the fact that, in a
representation in which configuration variables acted
by multiplication, the operator norm would coincide
with the sup-norm.} (i.e. the supremum norm), we
obtain a commutative $C^*$-algebra with identity.
Gel'fand theory ensures then that this algebra is
(isomorphic to) that of continuous functions on a
certain compact space, $\bar{A}$, which is usually
called the spectrum \cite{vel}. Smooth connections are
dense in this space $\bar{A}$ of quantum generalized
connections. Besides, the Hilbert space of any
representation of the $C^*$-configuration algebra is
of the form $L^2( \bar{A}, \mu)$ for some measure
$\mu$. The LOST theorem guarantees that there is a
unique Hilbert space $L^2( \bar{A}, \mu_{AL})$
supporting a representation not just of the
holonomies, but of the whole holonomy-flux algebra,
and such that $\mu_{AL}$ (the so-called the
Ashtekar-Lewandowski measure) is a
diffeomorphism-invariant, regular Borel measure. This
representation turns out not to be continuous and, as
an important consequence, the connection itself cannot
be defined as an operator valued distribution
\cite{ashlew}.

\section{Loop Quantum Cosmology: Flat FRW model}

LQC confronts the quantum analysis of cosmological
systems by applying similar quantization techniques to
those described for LQG. Here, we will focus our
discussion on a simple but physically relevant model,
namely, the case of homogeneous and isotropic flat
(FRW) cosmologies. As the matter content, we will
consider a massless minimally coupled scalar field.

The spatial manifold $\Sigma$ is topologically
$I\!\!R^3$, endowed with the action of the Euclidean
group. One can introduce a fiducial flat co-triad,
$\,^{0}e_a^i$, with the corresponding fiducial metric
and triad, $\,^{0}h_{ab}$ and $ \,^{0}e^a_i $. Given
the non-compactness of $\Sigma$, we also choose a
reference cell adapted to the fiducial triad in order
to integrate homogenous quantities, such as the
symplectic structure or the Hamiltonian, without
introducing infinities in our formalism. We use the
symbol $V_0$ to denote the fiducial volume of this
cell. Actually, physical results can be proven
independent of these choices under a suitable
definition of the elementary variables
\cite{aps,aps2}. In more detail, one can fix the gauge
and diffeomorphism freedom so that
\begin{equation} A_a= c \,V_0^{-1/3} \;\;^{0}e_a^i
\tau_i,\quad {\rm and} \quad E^a=p \, V_0^{-2/3}
\sqrt{{\rm det} \,^{0}h} \;\;^{0}e^a_i
\tau^i.\end{equation} Here, $c$ and $p$ are constant
on $\Sigma$ (but not under evolution), and describe
the only remaining degrees of freedom in our basic
variables. In the following, we call $\Gamma^{S}$ the
subspace of the gravitational phase space (for full
General Relativity) defined in this way.

The gravitational symplectic structure induced on
$\Sigma$ is just \begin{equation} \Omega^S= 3 {\bf d}
c \wedge {\bf d} p, \quad {\rm so} \;\; {\rm that}
\quad \left\{ c, p \right\}= \frac{1}{3}. \label{can}
\end{equation} The variables $c$ and $p$ are hence canonical,
apart from the factor of $1/3$. Holonomies along
straight edges $\mu \,^{0}e^a_i$ in the fiducial
directions suffice to separate symmetric connections,
i.e., given two different connections, there always
exists an edge of this kind for which the
corresponding holonomies differ \cite{vel}. We thus
restrict our attention to those holonomies,
$h_{\,^0e_i}(\mu)$, which have the form
\begin{equation} h_{\,^0e_i}(\mu)=\cos{\left(\frac{\mu
c}{2}\right)} {\bf 1} + 2 \sin{\left(\frac{\mu
c}{2}\right)} \tau_i.\end{equation} Similarly,
densitized triads can now be smeared just across
squares with edges parallel to the fiducial
directions,
\begin{equation} E(S,f)= p \;^{0}A(S,f)
V_0^{-2/3}.\end{equation} The factor $\;^{0}A(S,f)$
measures only the fiducial area of $S$ weighted with
an orientation factor. In this sense, fluxes are
totally determined by $p$, which therefore plays the
role of a momentum. The configuration algebra, on the
other hand, is generated by sums of products of matrix
elements of holonomies. Thus, it is the linear space
of continuous and bounded complex functions in
$I\!\!R$ provided by finite sums of the form $f(c)=
\sum_n f_n \;e^{i \mu_n c}$. Its completion with
respect to the sup-norm is known to be (isomorphic to)
the Bohr $C^*$-algebra of almost periodic functions
\cite{vel}.

\section{Bohr compactification and polymer representation}

As we have commented, the configuration $C^*$-algebra
is the algebra of almost periodic functions. The
(Gel'fand) spectrum of this algebra is the Bohr
compactification of the real line, $I\!\!R_{Bohr}$
\cite{vel}. This compactification can be understood as
the set of group homomorphisms from the group $I\!\!R$
(with the sum) to the multiplicative group $T$ of
complex numbers with unit norm. So, every $x\in
I\!\!R_{Bohr}$ is a map $x : I\!\!R \rightarrow T$
which satisfies
\begin{equation} x(0)=1,\quad\quad
x(p_1+p_2)=x(p_1)x(p_2)  \quad \forall p_1, p_2\in
I\!\!R .\end{equation} Since $T$ is a commutative
group, the operation $x \tilde x (p):= x(p) \tilde x
(p)$ provides a commutative group structure in
$I\!\!R_{Bohr}$. This group is compact with respect to
the Tychonoff product topology. We recall that the
Tychonoff topology is the weakest topology for which
the functions $F_p: I\!\!R_{Bohr} \rightarrow T$ given
by evaluation at $p$ [i.e. $F_p(x):= x(p)$] are all
continuous for any $p\in I\!\!R$ \cite{vel}. Besides,
the real line is actually dense in $I\!\!R_{Bohr}$.
This result follows from the fact that the algebra of
functions $f(c)$ considered at the end of the previous
section separates points $c\in I\!\!R$ \cite{vel}.

The compact group $I\!\!R_{Bohr}$ is equipped with a
normalized invariant measure under the group
operation, namely, the Haar measure $\mu_H$. The
representation of the holonomy-flux algebra for LQC is
given precisely by the Hilbert space
$L^2(I\!\!R_{Bohr}, \mu_H)$. In addition, since
$\mu_H$ is invariant under multiplication in the
group, we get that, $\forall \tilde{x} \in
I\!\!R_{Bohr}$,
\begin{equation}  [ 1- \tilde{x}(p)]
\int_{I\!\!R_{Bohr}} F_p(x) d\mu_H(x)
=0,\end{equation} from where it follows that
\begin{equation} \int _{I\!\!R_{Bohr}} F_p(x) d\mu_H(x)=
\delta_{p}^0.\end{equation} Taking into account that,
from our definitions, $F_{p_1}F_{p_2}= F_{p_1+p_2}$
and $F_p^{\ast}=F_ {-p}$, it is straightforward to
conclude that the set $\left\{ F_p, p \in I\!\!R
\right\}$ is orthonormal (hence, the Hilbert space
$L^2(I\!\!R_{Bohr}, \mu_H)$ is nonseparable). One can
also see that this set is dense \cite{vel}. As a
consequence, the Hilbert space $L^2(I\!\!R_{Bohr},
\mu_H)$ is isomorphic to the so-called ``polymer''
space of functions of $p\in I\!\!R$ that are square
integrable with respect to the discrete measure. The
isomorphism is given by $I: F_p \rightarrow \left| p
\right\rangle$ $\forall p \in I\!\!R$.

Employing then the orthonormal basis
\begin{equation}\left\{ \left| p \right\rangle; \, p \in
I\!\!R ,  \langle \tilde{p} | p \rangle =
\delta^{\tilde{p}}_p \right\},\end{equation} and
introducing the notation $N_{\mu}:= \exp{(i\mu c/2)}$,
the polymer ``momentum'' representation is determined
by the following action of the holonomy and flux
operators:
\begin{equation} \hat{p} \left| p \right\rangle=
\frac{p}{6} \left| p \right\rangle,\quad\quad
{\hat{N}}_{\mu}\left| p \right\rangle=\left| p + \mu
\right\rangle.\end{equation} In this representation,
states take the general form \begin{equation}\left|
\psi \right\rangle:= \sum_{p \in I\!\!R} \psi(p)
\left| p \right\rangle; \quad\quad \sum_{p \in I\!\!R}
\left| \psi(p) \right|^2 < \infty .\end{equation} Note
that normalizable states $\psi(p)$ differ from zero
only on a countable subset of the real line for the
label $p$, because the sequence $\{\psi(p)\}$ is
square summable. On the other hand, it is worth
noticing that the representation (of $N_{\mu}$) fails
to be continuous in $\mu$, as can be easily seen by
realizing that the state $\left| p \right\rangle$ is
always orthogonal to $\left| p+\mu \right\rangle$
regardless of the value of $\mu \neq 0$. Therefore,
the connection operator $\hat{c}$ is not well defined,
in total parallelism with the situation discussed for
LQG. The failure of continuity makes evident that the
representation is inequivalent to the standard
Schrödinger one of geometrodynamics \cite{geo} (often
called the Wheeler-DeWitt representation). This lack
of continuity explains why the Stone-von Neumann
uniqueness theorem of Quantum Mechanics does not apply
\cite{simon}, allowing the results of LQC to differ
radically from those --physically unsatisfactory--
attained in geometrodynamics.

\section{Quantum FRW model}

With our symmetry reduction to the flat FRW model and
our choice of fiducial structures, the triad adopts
the expression $e^a_i= {\rm sign}(p) | p |^{-1/2}
V_0^{1/3} \;^{0}e^a_i$. This triad diverges at the
big-bang singularity, corresponding to $p=0$. In the
quantum theory, on the other hand, $\hat{p}$ has just
a point spectrum \cite{gal} which coincides with the
whole real line, since the basis states $| p \rangle$
have unit norm $\forall p \in I\!\!R$. Since zero is
included in this point spectrum, the related (inverse)
operator ${| \hat{p} |}^{-1}$ is not well defined.
However, it is actually possible to define a triad
operator in terms of our elementary ones \cite{ABL}.
Classically, we have the following identity $\forall
\bar{\mu} \in I\!\!R$:
\begin{equation} \frac{{\rm sign}(p)}{\sqrt{ | p | }}
= \frac{4}{\bar{\mu}} {\rm tr} \left( \sum_{i=1}^3
\tau^i h_{\,^0e_i}(\bar{\mu}) \left\{ h_{\,^0e_i}^{-1}
(\bar{\mu}), \sqrt{ | p | } \right\} \right).
\end{equation} Here, $h_{\,^0e_i}$ is again the holonomy
along the edge $\,^0e_i$, and the symbol ${\rm tr}$
denotes the trace. Then, replacing Poisson brackets
with $-i$ times commutators, we obtain
\begin{equation} \frac{\bar{\mu}}{6}
\widehat{\left[\frac{{\rm sign}(p)} {\sqrt{ | p |} }
\right] } = {\hat{N}}_{-\bar{\mu}} | \hat{p} |^{1/2}
{\hat{N}}_{\bar{\mu}} - {\hat{N}}_{\bar{\mu}} |
\hat{p} |^{1/2} {\hat{N}}_{-\bar{\mu}} .\end{equation}
It is not difficult to check that this operator is
diagonal in the $p$-basis. Furthermore, it is bounded
from above, so that the classical divergence at $p=0$
disappears quantum mechanically with this
regularization of the triad \cite{ABL}. In fact, this
triad operator is such that it annihilates the state
$| p=0 \rangle$.

Since we have already fixed the gauge and
diffeomorphism freedom, the only constraint remaining
in the system is the Hamiltonian one. For flat FRW
spacetimes with a massless scalar field $\phi$ (and
unit lapse), this constraint can be obtained from the
evaluation of the following expression in the symmetry
reduced model \cite{aps,aps2}
\begin{equation}  H := \frac{1}{2}\int_{I\!\!R^3} | {\rm
det} E |^{-1/2} \left( P_{\phi}^2
-\epsilon_{\;\;\,k}^{ij}  E^a_i E^b_j F_{ab}^k
\right)=0 .\end{equation} To define the operator
corresponding to $| {\rm det} E |^{-1/2}$ (or to $|
{\rm det} E |^{-1/2} \epsilon_{\;\;\,k}^{ij} E^a_i
E^b_j$ in the gravitational part of the constraint
\cite{ABL}) we proceed as we have explained above when
discussing the triad operator. In addition, to
introduce an operator representation for the curvature
$F_{ab}^k$, we first recall the classical relation
\begin{equation} F_{ab}^k = -2 \lim_{
\bar{\mu}\rightarrow 0} {\rm tr} \left( \frac{h_{
[ij]}(\bar{\mu})-1 } {{\bar{\mu}}^2 V_0^{2/3}}
\tau^k\; \,^{0}e^i_a \,^{0}e^j_b \right
),\end{equation} which is valid for any real value of
$\bar{\mu}$ and where
\begin{equation} h_{ [ij]}(\bar{\mu}):=
h_{\,^0e_i}(\bar{\mu}) h_{\,^0e_j}(\bar{\mu})
h_{\,^0e_i}^{-1}(\bar{\mu})
h_{\,^0e_j}^{-1}(\bar{\mu}).\end{equation}
Nonetheless, after substituting classical holonomies
by their quantum counterparts, the limit of zero
regulator $\bar{\mu}$ cannot be taken in the resulting
curvature operator. This circumstance is interpreted
as a manifestation of the fact that, in LQG, the area
spectrum is discrete with a minimum nonzero eigenvalue
\cite{thie,area}, so that the square with edges
$\bar{\mu}\,^0e_i$ and $\bar{\mu}\,^0e_j$, employed to
define $h_{ [ij]}(\bar{\mu})$, cannot be shrunk to
zero. The regulator is then fixed by demanding that
the physical area of this square equals the minimum
nonvanishing eigenvalue allowed in LQG, which we call
$\Delta$ from now on. Hence, one gets the operator
relation ${\bar{\mu}}^2 | \hat{p} | = \Delta$.

At this stage, it is convenient to relabel the
$p$-basis by introducing the affine parameter
associated with the vector field
$\frac{1}{6}{\bar{\mu}}\partial_p$ \cite{aps2}. This
vector field can be regarded as that corresponding to
the exponent $\frac{1}{2}{\bar{\mu}c}$ in the holonomy
$N_{\bar{\mu}}$. Taking into account that the physical
volume of the fiducial cell is given by the operator
$\hat{V}= | \hat{p} |^{3/2}$, the above relabeling
leads to a basis of volume eigenstates $| \nu
\rangle$, where $\nu= 4 \;{\rm sign}(p) | p
|^{3/2}/\sqrt{\Delta}$. The operator
${\hat{N}}_{\bar{\mu}}$ is then defined to produce a
constant unit shift in the new label \cite{aps2},
\begin{equation} {\hat{N}}_{\bar{\mu}} \left| \nu
\right\rangle := \left| \nu +1 \right\rangle
.\end{equation}

Using the standard Schrödinger representation for the
matter field, so that the total Hilbert space is the
tensor product of the polymeric one and of
$L^2(I\!\!R, d\phi)$, and adopting a suitable factor
ordering, one finally arrives at the following quantum
Hamiltonian constraint:
\begin{eqnarray}
\hat{H} &:=& \frac{1}{2}{\widehat{ \left[
\frac{1}{\sqrt{ | p |}} \right]}}^{3/2} \left( -6
{\widehat{\Omega}}^2 + {\hat{P}}_{\phi}^2 \right)
{\widehat{\left[ \frac{1}
{\sqrt{ | p |}} \right]}}^{3/2},\\
{\widehat{\Omega}} &:=& \frac{1}{4\sqrt{\Delta}i}
\left[ \widehat{ \frac{1}{ \sqrt{|p|}} }
\right]^{-1/2} {\widehat{ {\sqrt{{|p|}}}  }}   \left[
\left({\hat{N}}_{2 \bar{\mu}}-{\hat{N}}_{-2 \bar{\mu}}
\right)  \widehat{ {\rm sign}{(p)}}+ \widehat{ {\rm
sign}{(p)}} \left({\hat{N}}_{2 \bar{\mu}}
-{\hat{N}}_{-2 \bar{\mu}} \right) \right]\nonumber\\
&\cdot& {\widehat{ {\sqrt{{|p|}}}  }} \left[ \widehat{
\frac{1}{ \sqrt{|p|}} } \right]^{-1/2}.
\end{eqnarray}
The symmetric factor ordering adopted for
${\widehat{\Omega}}$ arises naturally from the
consideration of homogeneous but anisotropic models of
Bianchi I type, where the ordering is well motivated,
regarding the flat FRW cosmologies as a special case
with vanishing anisotropies \cite{beni}. It is
straightforward to check that the above quantum
constraint annihilates the state $\left | p=0
\right\rangle$ (or equivalently $\left | \nu=0
\right\rangle$) and leaves invariant its orthogonal
complement. In the search for nontrivial solutions to
the constraint, one can then restrict all
considerations to this orthogonal complement, so that
the classical singularity, corresponding to $p=0$, can
be removed from the kinematical (gravitational)
Hilbert space \cite{beni,beni2}. In this sense, the
big-bang singularity is already resolved quantum
mechanically (see also \cite{boj}).

\section{Densitized constraint}

Once the state $\left | \nu=0 \right\rangle$ has been
removed, let us call $Cyl_{S}^{\neq}$ the linear span
of the nonzero volume eigenstates $\left\{ \left| \nu
\right\rangle ; \nu \neq 0, \nu \in I\!\!R \right\}$.
Based on previous experience with gravitational
models, we expect nontrivial solutions to the
constraint to live in the algebraic dual of
$Cyl_{S}^{\neq}$. Since the operator $\widehat{ [
{1}/{\sqrt{ | p |}}] }$ is invertible in the
orthogonal complement of $\left| \nu=0 \right\rangle$,
it is easy to check that one gets a bijection between
the considered solutions and those of the alternative
``densitized'' constraint \cite{beni}
\begin{equation} \hat{C} := -6 {\widehat{\Omega}}^2 +
{\hat{P}}_{\phi}^2 .\end{equation}

The operator ${\widehat{\Omega}}^2$ (with domain
$Cyl_{S}^{\neq}$) has the following action:
\begin{eqnarray} {\widehat{\Omega}}^2 \left| \nu
\right\rangle = &-& f_{+}(\nu)f_{+}(\nu+2) \left|
\nu+4 \right\rangle +
\left[f^2_+(\nu)+f^2_{-}(\nu)\right] \left| \nu
\right\rangle \nonumber \\ &-& f_{-}(\nu)f_{-}(\nu-2)
\left| \nu-4 \right\rangle ,\end{eqnarray} where
\begin{equation}
f_\pm(\nu)=\frac{1}{4\sqrt{6}\, \Delta^{1/4}}
g(\nu\pm2)s_\pm(\nu)g(\nu) , \quad\quad
s_\pm(\nu)={\rm sign}(\nu\pm 2)+{\rm sign}(\nu),
\end{equation}
and
\begin{equation}
g(\nu)=
\left|\left|1+\frac{1}{\nu}\right|^{\frac{1}{3}}
-\left|1-\frac{1}{\nu}\right| ^{\frac{1}{3}}
\right|^{-\frac{1}{2}}  \quad {\rm if} \quad \nu\neq
0, \end{equation} while $g(\nu=0)=0$. We notice that
${\widehat{\Omega}}^2$ relates only states $| \nu
\rangle$ whose label differ by a multiple of four.
Moreover, owing to the linear combination of signs in
$s_\pm(\nu)$, one can see that the real function
$f_+(\nu)f_{+}(\nu+2)$ has a remarkable property,
namely, it vanishes in the whole interval [-4,0].
Something similar happens with
$f_{-}(\nu)f_{-}(\nu-2)$, which vanishes in [0,4]. As
a consequence, for the label $\nu$, the action of the
operator ${\widehat{\Omega}}^2$ does not mix any of
the semilattices ${\mathcal L}^{\pm}_{\varepsilon}:=
\left\{ \pm (\varepsilon+4n), \; n \in I\!\!N
\right\}$, with $\varepsilon \in (0,4]$ --but
otherwise unspecified. In the following, we call ${\rm
H}^{\pm}_{\varepsilon}$ the corresponding Hilbert
subspaces of states with support in these semilattices
(i.e., the completion of the linear span of
$\nu$-states with $\nu\in {\mathcal
L}^{\pm}_{\varepsilon}$). Each of these subspaces can
be considered a superselection sector for the quantum
theory, inasmuch as they provide irreducible
representations for the physically relevant operators
of the model \cite{aps,aps2}.

On the other hand, it is possible to prove that (up to
a global multiplicative factor)
${\widehat{\Omega}}^2$, restricted to ${\rm
H}^+_{\varepsilon}\cup{\rm H}^-_{4-\varepsilon}$,
differs by a symmetric, trace-class operator from an
operator which is unitarily related with the
Hamiltonian of a point particle in a Pösch-Teller
potential \cite{beni,lewandgang}. From the properties
of this Hamiltonian and Kato perturbation theory
\cite{kato}, it then follows that
${\widehat{\Omega}}^2$ is essentially self-adjoint and
that its absolutely continuous spectrum\footnote{See,
e.g., Appendix C in reference \cite{gal} for a summary
on operator theory.} is $I\!\!R^+$. The rest of the
spectrum can be proven empty \cite{beni}. Moreover,
${\widehat{\Omega}}^2$ commutes with the projections
to ${\rm H}^+_{\varepsilon}$ and ${\rm
H}^-_{4-\varepsilon}$. One can then see that the
operator on any of these Hilbert spaces is positive
with an absolutely continuous spectrum of unit
degeneracy \cite{beni}.

Hence, on any superselection sector ${\rm
H}^{\pm}_{\varepsilon}$, one obtains a spectral
decomposition of the identity of the form
\begin{equation} {\bf {1}}_{\pm\epsilon} =
\int_{0}^{\infty} d\lambda | e_{\lambda}^{\pm\epsilon}
\rangle \langle e_{\lambda}^{\pm\epsilon}
|,\end{equation} where $| e_{\lambda}^{\pm\epsilon}
\rangle$ is a generalized eigenstate of
${\widehat{\Omega}}^2$ with eigenvalue equal to
$\lambda$. Finally, let us comment that, expressing $|
e_{\lambda}^{\pm\epsilon} \rangle$ in the $\nu$-basis,
the corresponding generalized eigenfunctions
$e^{\pm\epsilon}_{\lambda}(\nu)$ can always be chosen
real.

\section{Physical states}

Employing the above spectral decomposition associated
with ${\widehat{\Omega}}^2$, elements of the polymer
space ${\rm H}^{\pm}_{\varepsilon}$ can be identified
with elements of the Hilbert space $L^2(I\!\!R^{+},
d\lambda)$. It is now straightforward to solve the
densitized constraint $-6 {\widehat{\Omega}}^2 +
{\hat{P}}_{\phi}^2 =0$. Starting from the kinematical
Hilbert space ${\rm H}^{\pm}_{\varepsilon} \otimes
L^2(I\!\!R, d\phi)$, the solutions adopt the form
\begin{equation} \psi (\nu, \phi) =
\int_0^{\infty} d\lambda \;
e^{\pm\varepsilon}_{\lambda}(\nu) \left[
\psi_{+}(\lambda) e^{i  \sqrt{6\lambda} \phi} +
\psi_{-}(\lambda) e^{-i \sqrt{6\lambda} \phi} \right].
\label{phy}\end{equation}

Physical states can be identified with positive
frequency solutions, and hence with wavefunctions in
$L^2(I\!\!R^{+}, d\lambda)$. A complete set of Dirac
observables (acting on physical states) is given by
${\hat{P}}_{\phi}$ and, e.g., $| \hat{\nu}
|_{\phi_0}$, the latter being defined by the action of
$| \hat{\nu} |$ when $\phi=\phi_0$.
\footnote{Recalling expression (\ref{phy}), this
suffices to determine the action of the operator on
positive frequency solutions for all values of
$\phi$.} In this way, the LQC approach succeeds in
achieving a complete quantization of the flat FRW
model with massless scalar field.

Rather than in general physical states, one is usually
interested in states which display a semiclassical
behavior in the region of large spatial volumes and
matter fields, so that they can be regarded as
potential candidates to explain the properties of
universes like the one which we observe. With this
motivation, we can concentrate our considerations on
positive frequency states which, for a fixed large
value of the scalar field $\phi=\phi_0\gg 1$, are
peaked on certain values $P_{\phi}= P_{\phi}^0$ and
$\nu = \nu^0$ of the Dirac observables such that $|
\nu^0 |\gg 1$ and $ | P_{\phi}^0 | \gg 1$ \cite{aps2}.
In more detail, one analyzes Gaussians of the form
\begin{equation}
\psi_{+}\left( \lambda= \frac{\omega^2}{6} \right)
\propto e^{-{{(\omega+P_{\phi}^{0})^2}/{(2
\sigma^2)}}} e^{-i  \omega \phi_1} , \quad{\rm
with}\quad \phi_1=\phi_0 - \sqrt{\frac{2}{3}} \ln{
|\nu^0 |} .
\end{equation}
Numerical integration of the quantum evolution
dictated by the densitized Hamiltonian constraint
shows that the state remains peaked on a trajectory
which coincides with the union of a contracting and a
expanding classical solutions except in the region
where the matter energy density becomes comparable to
the Planck density in order of magnitude \cite{aps2}.
At that moment, the effective trajectory on which the
state is concentrated passes from a contraction to an
expansion phase in such a way that the classical
singularity is avoided. This quantum phenomenon which
allows the resolution of the big bang singularity is
usually called big bounce \cite{aps,aps2}.

\section{Conclusion}

We have seen that the quantization techniques of LQG
prove to be successful in achieving a rigorous and
complete quantum theory of simple cosmological models,
like e.g. the case of flat FRW spacetimes provided
with a minimally coupled scalar field. The resulting
quantization adopted in LQC is inequivalent to the
standard Schrödinger (or Wheeler-DeWitt) quantization
which has been traditionally employed in
geometrodynamics, a fact that explains why the
emerging physics is radically different and allows LQC
to supply satisfactory answers to fundamental problems
that had remained open in Quantum Cosmology. In
particular, this explains why, while the standard
quantization fails to solve the cosmological
singularities, these are cured in LQC. Actually, the
singularities are resolved already at the kinematical
level. Nonetheless, the resolution is much stronger.
For physical states with good semiclassical behavior,
numerical simulations show that the universe suffers a
big bounce before reaching the big bang. This bounce
occurs when the energy density $\rho= P_{\phi}^2/(2 |
p |^3)$ approaches a critical density of the order of
the Planck density. Away from the bounce, states are
peaked on classical solutions. Quantum corrections are
strong close to the bounce, but even there the state
remains peaked on a certain, modified trajectory.

\section*{Acknowledgments}

The author wants to thank the organizers of the XVII
IFWGP for the nice atmosphere they collaborated to
create during the workshop. He is in debt with J. M.
Velhinho for enlightening conversations and
discussions about the fundamentals of loop quantum
cosmology, as well as for explanations about technical
aspects of the polymer quantization, without which
this introduction would not have been possible. This
work was supported by the Spanish MEC Grant
FIS2005-05736-C03-02, its continuation
FIS2008-06078-C03-03, and the Spanish
Consolider-Ingenio 2010 Programme CPAN
(CSD2007-00042).


\begin{thebibliography}{29}

\bibitem{hawk} S. Hawking, and G. F. R. Ellis,
\emph{The Large Scale Structure of Space-Time},
Cambridge University Press, Cambridge (UK), 1973.

\bibitem{LQG} A. Ashtekar, \emph{Lectures on
Non-Perturbative Canonical Gravity}, edited by L. Z.
Fang, and R. Ruffini, World Scientific, Singapore,
1991.

\bibitem{LQG2} C. Rovelli, \emph{Quantum Gravity},
Cambridge University Press, Cambridge (UK), 2004.

\bibitem{thie} T. Thiemann, \emph{Modern Canonical
Quantum General Relativity}, Cambridge University
Press, Cambridge (UK), 2007.

\bibitem{LQC} M. Bojowald, \emph{Living Rev. Relativity}
\textbf{11}, 4 (2008).

\bibitem{ham} See, e.g., R. M. Wald,
\emph{General Relativity}, University of
Chicago Press, Chicago, 1984.

\bibitem{immirzi} G. Immirzi, \emph{Nucl. Phys. B
(Proc. Suppl.)} \textbf{57}, 65--72 (1997).

\bibitem{immirzi2} G. Immirzi, \emph{Class. Quantum Grav.}
\textbf{14}, L177--L181 (1997).

\bibitem{bhentropy} A. Ashtekar, J. C. Baez, and
K. Krasnov, \emph{Adv. Theor. Math. Phys.} \textbf{4},
1--94 (2001).

\bibitem{referee} J.N. Goldberg, J. Lewandowski, and C.
Stornaiolo, \emph{Commun. Math. Phys.} \textbf{148},
377-402 (1992).

\bibitem{referee2} A. Ashtekar and J. Lewandowski,
\emph{Class. Quantum Grav.} \textbf{9}, 1433-1468
(1992).

\bibitem{ashlew} A. Ashtekar, and J. Lewandowski,
\emph{Class. Quantum Grav.} \textbf{21}, R53--R152
(2004).

\bibitem{lost} J. Lewandowski, A. Okolow, H. Sahlmann,
and T. Thiemann, \emph{Comm. Math. Phys.}
\textbf{267}, 703-733 (2006).

\bibitem{vel} J. M. Velhinho, \emph{Class. Quantum
Grav.}, \textbf{24}, 3745-3758 (2007).

\bibitem{aps} A. Ashtekar, T. Pawlowski, and P.
Singh, \emph{Phys. Rev. D} \textbf{73},
124038/1--124038/33 (2006).

\bibitem{aps2} A. Ashtekar, T. Pawlowski, and P.
Singh, \emph{Phys. Rev. D} \textbf{74},
084003/1--084003/23 (2006).

\bibitem{simon} B. Simon, \emph{Topics in Functional
Analysis}, edited by R. F. Streater, Academic Press,
London, 1972.

\bibitem{gal} See, e.g., A. Galindo, and P. Pascual,
\emph{Quantum Mechanics I}, Springer-Verlag, Berlin,
1990.

\bibitem{geo} See, e.g., J. J. Halliwell, ``Introductory
Lectures on Quantum Cosmology'', in \emph{Proceedings
of the Seventh Jerusalem Winter School for Theoretical
Physics: Quantum Cosmology and Baby Universes}, edited
by S. Coleman, J. B. Hartle, T. Piran, and S.
Weinberg, World Scientific, Singapore, 1991, pp.
159--243.

\bibitem{ABL} A. Ashtekar, M. Bojowald, and J.
Lewandowski, \emph{Adv. Theor. Math. Phys.}
\textbf{7}, 233--268 (2003).

\bibitem{area} A. Ashtekar, and J. Lewandowsky,
\emph{Class. Quantum Grav.} \textbf{14}, A55-A81
(1997).

\bibitem{beni} M. Mart\'{\i}n-Benito, G. A. Mena
Marug\'{a}n, and T. Pawlowski, \emph{Phys. Rev. D}
\textbf{78}, 064008/1--064008/11 (2008).

\bibitem{beni2} M. Mart\'{\i}n-Benito, G. A. Mena
Marug\'{a}n, and L. Garay, \emph{Phys. Rev. D}
\textbf{78}, 083516/1--083516/5 (2008).

\bibitem{boj} M. Bojowald, \emph{Class. Quantum Grav.}
\textbf{20}, 2595--2615 (2003).

\bibitem{lewandgang} W. Kami\'nski, and J. Lewandowski,
\emph{Class. Quantum Grav.} \textbf{25},
035001/1--035001/11 (2008).

\bibitem{kato} T. Kato, \emph{Perturbation Theory for
Linear Operators}, Springer-Verlag, Berlin, 1980.

\end{thebibliography}
\end{document}